\documentclass[pra,showpacs]{revtex4}
\usepackage{graphicx}
\usepackage{amssymb}
\long\def\ca#1\cb{}

\def\dyad#1#2{|#1\rangle\langle#2|}

\def\ket#1{|#1\rangle }

\def\mted#1#2#3{\langle#1|#2|#3\rangle }

\def\DC{{\cal D}}

\def\HC{{\cal H}}

\def\SC{{\cal S}}
\def\TC{{\cal T}}
\def\UC{{\cal U}}
\def\VC{{\cal V}}
\def\WC{{\cal W}}
\def\MC{{\cal M}}

\def\BC{{\cal B}}

\newtheorem{thm_block}{Theorem}
\newtheorem{lemma_block}[thm_block]{Lemma}
\newtheorem{pirrep}[thm_block]{Lemma}

\begin{document}
\title{Optimizing local protocols implementing nonlocal quantum gates}
\author{Scott M. Cohen$^{1,2}$}
\email{cohensm@duq.edu}
\affiliation{$^1$Department of Physics, Duquesne University, Pittsburgh,
Pennsylvania 15282\\
$^2$Department of Physics, Carnegie-Mellon University,
Pittsburgh, Pennsylvania 15213}

\begin{abstract}
We present a method of optimizing recently designed protocols for implementing an arbitrary nonlocal unitary gate acting on a bipartite system. These protocols use only local operations and classical communication with the assistance of entanglement, and are deterministic while also being ``one-shot", in that they use only one copy of an entangled resource state. The optimization is in the sense of minimizing the amount of entanglement used, and it is often the case that less entanglement is needed than with an alternative protocol using two-way teleportation.
\end{abstract}

\date{Version of 10 July 2009}
\pacs{03.67.Ac}

\maketitle
\section{Introduction}
Recent years have seen enormous advances in the study of quantum information, including in specific areas such as quantum computing \cite{NielsenChuang,ShorFactor,Shor_expand,Grover}. One particular proposal for a quantum computer attempts to overcome the difficulty of protecting a large quantum processor from decoherence by using a multi-processor design so that the size of each processor may be kept small. Obviously, such a ``distributed" quantum computer \cite{CiracDistComp} will require the implementation of tasks that involve multiple processors, including information transfer, measurements, and other nonlocal operations utilizing unitary gates. One possibility for performing such operations is to physically move component quantum systems around, but it may well be advantageous to perform these operations without bringing parts of different processors together, but rather by utilizing the power of entanglement. In such a situation, one is considering spatially separated parties restricted to local quantum operations on their individual subsystems with communication of classical information amongst the parties (LOCC). Although numerous tasks exist that remain possible under the LOCC paradigm when entanglement is not available --- an important example involves distinguishing between two orthogonal quantum states \cite{Walgate} --- there are also many important cases where global operations are required. However, even the latter tasks become possible with LOCC when there is enough entanglement available to the parties. This is one of the main reasons why entanglement is now recognized to be a valuable resource, especially when considering that it is difficult to create and maintain. An important question, then, is this: For a given nonlocal task, how much entanglement is necessary?

Due to the importance of such questions, much attention has been focused on the design and optimization of protocols utilizing entanglement to perform nonlocal tasks \cite{BennettEntPRL,TerhalDense,BennettUcaps,CiracEnt,DBerry,WoottersMeasure,EisertNLU,ReznikStator,MyLDUPB,OurNLU}. Such protocols fall into a variety of classes, including those that utilize an asymptotically large number of inputs in order to obtain a fraction of outputs having vanishingly small error \cite{BennettEntPRL,TerhalDense}. In contrast, deterministic protocols require perfect success with each individual input. In this case, one may consider that the given task will be repeated many times with the goal to optimize entanglement resources in terms of the average amount used over all repetitions \cite{DBerry,WoottersMeasure}. On the other hand, one may also consider one-shot protocols \cite{EisertNLU,ReznikStator,MyLDUPB,OurNLU}, in which the task is only performed once. Then, one wishes to know the smallest amount of entanglement with which the task may be accomplished in a single implementation. While these latter protocols will generally require the most entanglement, they do offer the advantage of certainty (which may at times be needed) and as such a savings in other resources, including the amount of time needed for the task.

Here, I consider the one-shot deterministic implementation of nonlocal unitaries acting on bipartite system $\HC_A\otimes\HC_B$ using only LOCC with the assistance of entanglement. We have recently succeeded in creating efficient protocols, with the most general of these having the ability to implement any bipartite unitary \cite{OurNLU}. The crucial remaining question is to determine the level of efficiency, in terms of the amount of entanglement that is required, that can be attained with these protocols. One way to measure efficiency is by comparison to a ``standard" protocol, in which the state of one subsystem, say $\HC_A$ of dimension $D_A$ (let us assume $D_A\le D_B$), is teleported to the laboratory of the other party, who performs the desired nonlocal unitary on the combined systems before teleporting the final state of the first party's subsystem back to where it belongs. This standard protocol requires two maximally entangled states (MES), each of rank $D_A$, so uses an amount of entanglement equal to $2\log D_A$. We thus wish to know under what circumstances, and in particular for which nonlocal unitaries, our protocol succeeds while using less entanglement than this. This question was essentially answered in \cite{OurNLU}, where a method was provided that parameterizes all bipartite unitaries that can be implemented by our protocol using a given amount of entanglement. However, it was not known how to determine the minimum amount needed for a particular, given unitary, a question of crucial importance since circumstances will often dictate which specific unitary is needed. The purpose of the present paper is to provide an answer to this question.

The protocol of \cite{OurNLU} for implementing the nonlocal unitary $\UC$ is based on an expansion of the form
\begin{eqnarray}\label{UC}
	\UC = \sum_{f\in G} U(f)\otimes W(f),
\end{eqnarray}
where matrices $U(f)$, which act on $\HC_A$, make up a projective unitary representation of a finite group $G$. The $W(f)$ are arbitrary matrices, apart from the requirement that $\UC$ be unitary, and act on $\HC_B$. To deterministically implement this unitary, our protocol uses --- and in fact when $G$ is the smallest group allowing an expansion of $\UC$ in the form (\ref{UC}), requires \cite{myMES} --- an amount of entanglement equal to $\log |G|$, where $|G|$ is the order of the group $G$. Given $\UC$, it is always possible to arrive at such an expansion by taking the $U(f)$ to be the generalized Pauli operators of dimension $D_A$ (for a definition, see Eq.~(11) of \cite{OurNLU}). However, since the order of this group is $D_A^2$, this particular expansion uses the same amount of entanglement as does the standard protocol, so it will not generally be optimally efficient. The reason this group may always be used is that the given representation in terms of generalized Pauli matrices forms a basis of the space of operators acting on $\HC_A$. This will obviously not be true if one chooses a smaller group, and for this reason an arbitrary group will not generally allow an expansion of a particular, chosen $\UC$. So given an arbitrary $\UC$, we seek an ``optimal" expansion; that is, an expansion in terms of the smallest group possible. In the remainder of this paper, we will see how such an expansion may be found.

We begin in the following section with a description of the protocol of \cite{OurNLU}. Then, in Section~\ref{findopt}, we show how to find the optimal expansion for an arbitrary $\UC$, which involves choosing the smallest group $G$. We conclude with a discussion of our results in Section~\ref{conclude}.

\section{The protocol}
The protocol we consider in this paper has been extensively discussed in \cite{OurNLU}. Here, we give only a brief description. It is based on the use of a finite group
$G$, which may or may not be Abelian, having elements $f$, $g$, etc., with $e$ the identity. Group multiplication is indicated by $fg$.  The unitary operators $U(f)$ of (\ref{UC}) form a \emph{projective representation} of $G$ in the sense that
\begin{equation}
 U(f) U(g) = \mu(f,g) U(fg)
\label{muFG}
\end{equation}
for all $f$ and $g$ in $G$. Here the $\mu(f,g)$ are nonzero complex numbers
constituting a \emph{factor system}; in our case they have unit magnitude because
the $U(f)$ are unitary. Without loss of generality, one can always take $\mu(f,g)=1$ whenever $f$ or $g$ is equal to $e$ or when $g=f^{-1}$, and we will follow this convention throughout. In addition, associativity of group multiplication requires that
\begin{eqnarray}
	\mu(f,g)\mu(fg,h)=\mu(f,gh)\mu(g,h).
\end{eqnarray}

\begin{figure}[ht]
\includegraphics[scale=1]{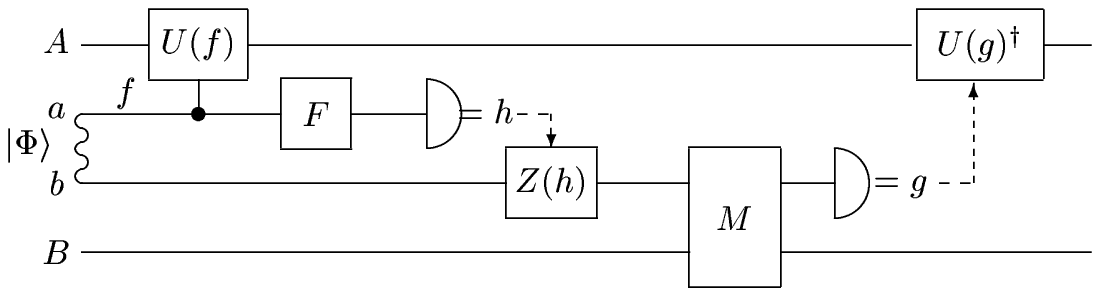}
\caption{Circuit diagram illustrating local implementation of
  nonlocal unitary ${\UC}=\sum_{f\in G}U(f)\otimes W(f)$, where the set of
  unitaries $\{U(f)\}$ forms a projective representation of a
  group.} \label{fig:bipartM}
\end{figure}

A circuit diagram for the protocol is given in Figure~\ref{fig:bipartM}. Alice and Bob share an MES,
\begin{eqnarray}
	|\Phi\rangle=\frac{1}{|G|}\sum_{f\in G}|f\rangle_a|f\rangle_b,
\end{eqnarray}
of Schmidt rank $|G|$ on systems $a,b$ described by Hilbert spaces $\HC_a,\HC_b$. Alice begins the protocol by implementing a controlled set of unitaries, where system $a$ in standard basis state $|f\rangle_a$ controls the unitary $U(f)$, which operates on $\HC_A$. She then measures $a$ in a basis unbiased to the standard basis and defined by unitary operator $F$, where by unbiased we mean that when represented in the standard basis, $F$ must have all entries of magnitude $|G|^{-1/2}$. Next, Alice tells Bob the outcome $h$ of that measurement, and he follows by performing diagonal unitary $Z(h)$ on $b$, which has the effect of removing phase factors introduced by Alice's measurement. He then performs unitary $M$ on $bB$, which introduces the operators $W(f)$ of (\ref{UC}) in a way that is correlated with the $U(f)$'s of Alice's earlier operation, where these correlations are made possible by the initial entanglement between $a$ and $b$. Bob then measures $b$ in the standard basis and tells Alice his outcome, $g$. Alice completes the protocol with the ``correction" $U(g)^\dagger$ on $A$, which adjusts the correlation between $W$'s and $U$'s so that the result is always equal to $\UC$ (that is, so that $W(f)$ is always tensored with $U(f)$, rather than with $U(f^\prime)$ for some $f^\prime\ne f$).

The matrix $M$ is given by
\begin{equation}
\label{eqq20}
M = \sum_{f\in G} R(f)\otimes W(f)
\end{equation}
with
\begin{equation}
\label{eqq19}
R(f) = \sum_{g\in G} \mu(g,f)\; \dyad{g}{gf},
\end{equation}
and these $R(f)$ form a projective regular representation of $G$ with factor system $\mu(f,g)$. It can be shown \cite{OurNLU} for the circuit of Figure~\ref{fig:bipartM} that the unitarity of $M$ implies that of $\UC$. By using the basis $\{\ket{g}\}$ of $\HC_b$ one can view $M$ as a matrix with blocks, where the $\mted{g}{M}{f}$ block is equal to $W(g^{-1}f)$ multiplied by a phase $\mu(g,g^{-1}f)$. This structure is responsible for the effectiveness of the correction $U(g)^\dagger$ in completing the protocol so that it successfully implements $\UC$ for every measurement outcome $g$ by Bob (and $h$ by Alice).

The reader is encouraged to consult \cite{OurNLU} for more details and further discussion.

\section{Finding the optimal expansion}
\label{findopt}
In this section, we show how to find an optimal expansion of the form (\ref{UC}) for an arbitrary unitary $\UC$; that is, an expansion in terms of a unitary representation $U(f)$ of the smallest possible finite group $G$. Recalling that the point of finding the \textit{smallest} group is to minimize the amount of entanglement required to implement $\UC$ using the protocol described in the previous section, we will consider local unitaries to be ``free". That is, instead of (\ref{UC}), we wish to consider the more general case of an expansion in the form,
\begin{equation}
\UC = (V_A^\prime\otimes V_B^\prime)\left(\sum_{f\in G} U(f)\otimes W(f)\right)(V_A\otimes V_B),
\end{equation}
with unitary $V_{A(B)}$ and $V_{A(B)}^\prime$. For a given group $G$, there will be unitaries that can be expanded in this way but not in the form (\ref{UC}). The local unitaries acting on $\HC_B$ simply lead to a replacement, $W(f)\rightarrow V_B^\prime W(f)V_B$, so these may be absorbed into the definition of $W(f)$ and effectively ignored. In addition,
\begin{equation}
V_A^\prime U(f)V_A=V_A^\prime V_A\left(V_A^\dagger U(f)V_A\right),
\end{equation}
and since $V_A^\dagger U(f)V_A$ is a representation of $G$ equivalent to the representation by $U(f)$, we may replace $U(f)\rightarrow V_A^\dagger U(f)V_A$. Thus, defining $V=V_A^\prime V_A$, we will include all possibilities by looking for an expansion of $\UC$ in the form,
\begin{equation}
\UC = \sum_{f\in G} [VU(f)]\otimes W(f).
\label{UClocal}
\end{equation}
[Note that, by the above arguments, it would be equally valid to place the $V$ to the right of the $U(f)$ in this expression.]

Therefore, given an arbitrary bipartite unitary $\UC$, we now see that our task of finding an optimal expansion requires the identification of: (1) the smallest group, $G$; (2) the projective representation, $U(f)$; (3) the local unitary, $V$; and finally, (4) the set of operators $W(f)$ acting on $\HC_B$. It should be clear that if $G$ is unknown, there is little hope of identifying the other items in this list. By first developing a method of finding $G$, we will see that the others follow without too much additional difficulty.

We note that one will also want to look at expanding in terms of a group on the $\HC_B$ side, as this may lead to a smaller group than the best expansion on the $\HC_A$ side even when $D_A<D_B$. Of course, if one finds an expansion with a group of order equal to the operator Schmidt rank of $\UC$, which is the number of terms in its Schmidt expansion (see Eq.(\ref{UCSchmidt}), below), then one is done since (by definition) no expansion can have a smaller number of terms than this.

\subsection{The optimal expansion of $\UC$}\label{opt}
We begin with the observation that every group is characterized by the dimensions of its irreducible representations (henceforth referred to as irreps). Our approach will then be to identify the set of irrep dimensions the group must contain, and then to find the smallest group $G$ that contains those dimensions. We know from group representation theory that the set $\{U(f)\}$ in (\ref{UClocal}) can be reduced --- that is, simultaneously brought to block-diagonal form --- by a unitary similarity transformation, and that the resulting blocks will each be associated with one of the irreps of $G$. Let us now see how we may identify the necessary irreps.

In order to make progress, we first need a way of expressing $\UC$. Let us use its operator Schmidt decomposition,
\begin{equation}\label{UCSchmidt}
\UC = \sum_j A_j\otimes B_j,
\end{equation}
where $\textrm{Tr}(B_k^\dagger B_j)=\delta_{jk}$, and  $\textrm{Tr}(A_k^\dagger A_j)=0~\forall{j\ne k}$. Given $\UC$, we can always find the $A_j$ and $B_j$, so we may assume these matrices are known. Equating the expansions in (\ref{UClocal}) and (\ref{UCSchmidt}), multiplying by $B_k^\dagger$, and tracing over $\HC_B$, we find that
\begin{equation}\label{AvsU}
V^\dagger A_k = \sum_{f\in G}\WC_{kf}U(f),
\end{equation}
with $\WC_{kf}=\textrm{Tr}(B_k^\dagger W(f))$. 

If we express (\ref{AvsU}) in the appropriate basis the $U(f)$ will all be block diagonal, and then so must be the product, $V^\dagger A_k$, for every $k$. If $V$ was known, then we could block-diagonalize the set of (known) matrices, $\{V^\dagger A_k\}$, and by relating these blocks to those of the $U(f)$ in (\ref{AvsU}), we could identify the size of the irreps that must be present in any group that the $U(f)$ could possibly represent. However, it is the $A_k$ that are known to us, $V$ and $U(f)$ being what we are trying to find. This means that (\ref{AvsU}) is not directly useful for identifying the irreps $G$ must contain. What we seek is a \textit{known} set of operators that can themselves be block-diagonalized, and at the same time expressed as a linear combination of the $U(f)$'s. The problem in the previous equation resides in the presence of $V$, which is unknown, so we will use a little trick to get rid of this operator.

Consider, then, multiplying (\ref{AvsU}) on the left by its Hermitian conjugate for index $j$, yielding
\begin{equation}\label{AdagA}
A_j^\dagger A_k = \sum_{f,g\in G}\mu(g^{-1},f)\WC_{jg}^\ast \WC_{kf}U(g^{-1}f),
\end{equation}
where use has been made of (\ref{muFG}). The objects on the left-hand side of this equation are known, and if the $U(f)$ are block-diagonal, then the $A_j^\dagger A_k$ will be so, as well. In fact, the above equation tells us that the \emph{finest block-diagonal form} of the set $\{U(f)\}$ cannot be finer than that of the set $\{A_j^\dagger A_k\}$. If we can find the finest block-diagonal form of $\{A_j^\dagger A_k\}$ and then choose the smallest $G$ that contains irreps of the same size as those resulting blocks, this will be the smallest $G$ that allows an expansion of the form (\ref{AdagA}). However, there is one potential problem with this approach, indicated by (\ref{AvsU}). This equation tells us that the finest block-diagonal form of $\{U(f)\}$ also cannot be finer than that of the set $\{V^\dagger A_k\}$, and it is not immediately obvious that the just-described method of choosing $G$ will necessarily satisfy this additional constraint. That is to say, it may be that there is no $V$ such that the set $\{V^\dagger A_k\}$ will have as fine a block-diagonal form as the set $\{A_j^\dagger A_k\}$. However, the following theorem tells us that for sets of operators of the type we are considering, this is not a problem.
\begin{thm_block}\label{th:block}
Given a set of operators $\{A_k\}$ taken from an expansion of a non-singular operator in the form $\sum_kA_k\otimes B_k$, there always exists a unitary $V$ such that the set $\{V^\dagger A_k\}$ has the same finest block diagonal form as the set $\{A_j^\dagger A_k\}$.
\end{thm_block}
The proof of this theorem, which also provides a method of constructing $V$, is given in Appendix~\ref{app:block}.

We therefore need to find the finest block-diagonal form of the $A_j^\dagger A_k$. It turns out that a very efficient method has recently been devised that accomplishes this task \cite{blockdiag2,blockdiag1}, providing the sizes of the necessary irreps that $G$ must contain, as well as the unitary $\SC$ that block-diagonalizes these operators. Thus, given $\UC$, we can find the size of the irreps that $G$ must have in order that an expansion of the form (\ref{UClocal}) is possible. Note that the number of different irreps needed is equal to the number of \textit{inequivalent} blocks in the $A_j^\dagger A_k$. For example, suppose that for every $j,k$, the finest block-diagonal form is
\begin{eqnarray}
	A_j^\dagger A_k = \left(\begin{array}{c c}
		 R_{jk}^{(1)} & 0\\
		 0 & R_{jk}^{(2)} \\
		\end{array}\right),
\end{eqnarray}
where the two blocks are of equal size $d$ and $R_{jk}^{(1)}=\TC R_{jk}^{(2)}\TC^\dagger$ for some fixed unitary $\TC$. Then these two blocks are ``equivalent" to each other, and there is only a single irrep needed in $G$, of size equal to $d$ and repeated twice in each $U(f)$ \cite{HermEqBlocks}.

An important implication of the proof of theorem \ref{th:block} is that the unitary $\SC$ which block-diagonalizes the $A_j^\dagger A_k$, and which is given by the method of \cite{blockdiag2,blockdiag1}, will also block-diagonalize the $V^\dagger A_k$. Then, once the group $G$ is chosen (see below), we will define the representation $U(f)$ to be block diagonal in the same basis as are the $V^\dagger A_k$ with the appropriate irreps in the appropriate positions. Then, since the proof of theorem \ref{th:block} provides a construction of $V$, we will have identified $V$ and the representation matrices $U(f)$. What is left is to find the operators $W(f)$, and then we will return to the question of actually choosing $G$.

To find $W(f)$, we note that
\begin{eqnarray}
	\UC &=& \sum_jA_j\otimes B_j = \sum_j\left(\sum_{f\in G}\WC_{jf}VU(f)\right)\otimes B_j\nonumber\\
		&=&\sum_{f\in G}[VU(f)]\otimes\left(\sum_j\WC_{jf}B_j\right)\nonumber\\
		&=&\sum_{f\in G}[VU(f)]\otimes W(f),
\end{eqnarray}
where in the first line we used (\ref{AvsU}) and the last line is just (\ref{UClocal}), allowing us to identify the choice,
\begin{eqnarray}\label{eq:W}
W(f)=\sum_j\WC_{jf}B_j. 
\end{eqnarray}
Matrix $\WC$ may be found as follows. Label the inequivalent blocks of $V^\dagger A_k$ by the irrep each is associated with, so that by (\ref{AvsU}),
\begin{eqnarray}\label{AvsWC}
	\left(V^\dagger A_k\right)^{(\lambda)} = \sum_{f\in G}\WC_{kf}U^{(\lambda)}(f),
\end{eqnarray}
with $U^{(\lambda)}(f)$ the $\lambda^{{th}}$ irrep of $G$. Define
\begin{eqnarray}\label{WC}
	\WC_{jg}=\sum_{\lambda=1}^\kappa \frac{d_\lambda}{|G|}\textrm{Tr}\left[U^{(\lambda)}(g^{-1})\left(V^\dagger A_j\right)^{(\lambda)}\right],
\end{eqnarray}
where the sum is over all the $\kappa$ inequivalent irreps belonging to $G$. By the well-known orthogonality condition of irreps \cite{Schensted}, 
\begin{eqnarray}
	\frac{|G|}{d_\lambda}~\delta_{\lambda\lambda^\prime}\delta_{mm^\prime}\delta_{nn^\prime}=\sum_{f\in G}\left(U^{(\lambda^\prime)}(f^{-1})\right)_{n^\prime m^\prime}\left(U^{(\lambda)}(f)\right)_{m n},
\end{eqnarray}
we see that this definition of $\WC$ automatically satisfies (\ref{AvsWC}) and if every irrep of $G$ appears in the representation by the $U(f)$, we are finished apart from the actual choice of $G$. If, on the other hand, any irreps $\lambda$ of $G$ are missing in the $U(f)$, $V^\dagger A_k$ will have no block $(V^\dagger A_k)^{(\lambda)}$ associated with these irreps, and for the purpose of calculating $\WC$ from (\ref{WC}), one may define the latter quantities in any way one wishes, including setting them to zero. This is consistent with the fact that missing irreps correspond to linearly dependent representations \cite{OurNLU}, and therefore a certain amount of freedom in the expansion coefficients $W(f)$ (compare (\ref{eq:W})). Note that given any set of $W(f)$'s, then by theorem $5$ of \cite{OurNLU} there is always at least one choice of coefficients for which the protocol described there will work. Thus, given any expansion in terms of the $U(f)$, such as one with the $W(f)$ obtained by setting $(V^\dagger A_k)^{(\lambda)}$ to zero in (\ref{WC}) for all the missing irreps, a method of choosing a workable set (that is, one for which our protocol will work) from the original set of $W(f)$'s in the linearly dependent case is described at the end of Appendix D in \cite{OurNLU}. Hence, we have shown how to obtain an expansion of any $\UC$ in terms of a group, an expansion that is optimal for our protocol once the smallest group $G$ is chosen. Let us now turn to the task of finding the smallest group.

\subsection{Choosing the smallest group}\label{chooseG}
We are left with the task of identifying the smallest group $G$ that contains a set of irreps of given dimensions, $\{d_\lambda\}$, where $\lambda$ labels the required inequivalent irreps (for convenience, order these as $d_1\le d_2 \le \cdots$). Here, I will describe how this can be done if confining oneself to ordinary representations; the (difficult, but probably often necessary) task of including projective representations will be discussed in Appendix~\ref{app:proj}. Now, since the order of a group, $|G|$, is equal to the sum of squares of the dimensions of its inequivalent irreps, it must be that $|G|\ge N_0\equiv\sum_\lambda d_\lambda^2$, where $\sum_\lambda$ here represents a sum over only those inequivalent irreps we have found are needed in $G$. Note also that $d_\lambda=\chi^{(\lambda)}(e)$; that is, the irrep dimensions are equal to the character of the identity element in that irrep. So, one approach would be to search through character tables of groups of order $N_0$, looking at characters of the identity element in the various irreps for these groups. If there exists such a group with characters of the identity element matching the desired set $\{d_\lambda\}$, then we are done and that group can be used for an optimal expansion of $\UC$. Otherwise, we need to look at groups of higher order $N>N_0$, but recalling that the irrep dimensions are divisors of $|G|$ (a fact that holds even for projective irreps \cite{Schur}), then given $\{d_\lambda\}$, this provides a strong constraint on possible values of $|G|$. So considering this constraint we can search through groups of higher and higher order until we find one containing all the required irreps. However, if before finding such a group we reach a value of $N=(d_1+d_2)^2+\sum_{\lambda>2}d_\lambda^2$, then one must also look for groups that have irrep dimensions, $d_1+d_2,d_3,\cdots$, along with looking for those with $d_1,d_2,\cdots$. The reason is that the two blocks of dimensions $d_1$ and $d_2$ may be considered as a single block of dimension $d_1+d_2$, with this larger block associated with a single irrep. This procedure may be continued to higher order groups, combining blocks as one proceeds (of size $d_1+d_3$, for example, and so on) until one finds a suitable candidate. However, if one reaches $N=D^2_A$ before finding a suitable group, one may conclude that it is not possible using the protocol of Figure~\ref{fig:bipartM} to implement $\UC$ more efficiently than by teleporting both directions, unless one can find a suitable projective representation of a smaller group, as described in Appendix~\ref{app:proj}. (Of course, one will also want to check to see if a smaller group can be found when the roles of Alice and Bob are reversed, the group representation matrices $U(f)$ then appearing on the $\HC_B$ side instead of $\HC_A$.)

[Another possible approach is the following. For each group satisfying the constraints mentioned in the previous paragraph, starting once again with the smallest, construct the regular representation in its standard form (permutation matrices with each row and column having only a single non-zero entry equal to one). This can be done in a straightforward way from the multiplication table of the group. Then the method of \cite{blockdiag2,blockdiag1} may be used to decompose this set of matrices into irreps, in the same way as was done above for the $A_j^\dagger A_k$. Note that it is only necessary to construct the regular representation for a set of generators of the group and then to decompose these, since if and only if the generators are in their finest block diagonal form, then the whole group will be so as well. This is an efficient method of finding not only the dimensions but also the irreps themselves, which will eventually be needed so that we can construct the $U(f)$. However, if a more efficient method (perhaps, for example, the one described in the previous paragraph) is able to determine the dimensions of the irreps of a group, this would be preferable, as we only need to know the irrep matrices of the group we ultimately choose to use. The irrep matrices for those groups that are rejected in the process are of no interest to us.]

These searches may be computationally demanding for large dimension, especially when looking for projective representations as described in Appendix~\ref{app:proj}, but will certainly be tractable if $D_A$ is not too large. We make no claim that the above methods (or that described in Appendix~\ref{app:proj} for projective representations) are optimally efficient, and we leave the question of optimizing this search as an open problem.

\subsection{Starting from expansions other than the Schmidt expansion}
There is one final issue that must be addressed. The arguments of the previous section are based on a very particular starting point, the Schmidt expansion of $\UC$. We must consider whether using a different starting point could lead to an expansion in terms of a smaller group. The answer is no, as the following discussion makes clear.

Suppose we start from a different expansion,
\begin{equation}\label{notSchmidt}
\UC = \sum_m \widetilde A_m\otimes \widetilde B_m.
\end{equation}
Equating this to (\ref{UCSchmidt}), the Schmidt expansion of $\UC$, multiplying by $B_k^\dagger$ and tracing over $\HC_B$, we find
\begin{equation}
A_k = \sum_m \beta_{km}\widetilde A_m,
\end{equation}
with $\beta_{km} = \textrm{Tr}(B_k^\dagger\widetilde B_m)$, and
\begin{equation}
A_j^\dagger A_k = \sum_{mn} \beta_{km}\beta_{jn}^\ast\widetilde A_n^\dagger\widetilde A_m.
\end{equation}
This tells us that the finest block-diagonal form of the set $\{\widetilde A_n^\dagger\widetilde A_m\}$ cannot be finer than the finest block-diagonal form of the set $\{A_j^\dagger A_k\}$. We may therefore conclude that no other starting expansion can lead to a finer block-diagonal form than when starting from the Schmidt expansion of $\UC$, and therefore the group obtained by the arguments of the previous section is in fact the smallest group possible. Indeed, a moment's thought leads to the conclusion that any expansion of the form (\ref{notSchmidt}) with linearly independent operators, $\{\widetilde B_m\}$, will lead to the \textit{same} finest block diagonal form as does the Schmidt expansion, and hence also to the smallest group (this could be useful, as it may be easier to obtain some other expansion, such as in terms of generalized Pauli operators, than to obtain the Schmidt expansion). The reason is that when the $\widetilde B_m$ are linearly independent, one may also obtain $\widetilde A_m$ as a linear combination of the $A_k$, so that the finest block-diagonal form of the set $\{A_j^\dagger A_k\}$ also cannot be finer than the finest block-diagonal form of the set $\{\widetilde A_n^\dagger\widetilde A_m\}$. The conclusion follows immediately, and it is therefore acceptable to use any starting expansion one chooses, as long as the $\{\widetilde B_m\}$ are a linearly independent set.

\section{Conclusions}\label{conclude}
We have presented a detailed method for finding an optimal expansion of a nonlocal unitary $\UC$, allowing $\UC$ to be implemented with the smallest amount of entanglement possible when using the protocol of \cite{OurNLU}. This method utilizes an expansion of $\UC$ in terms of a projective representation of a finite group $G$, where the order (size) of $G$ determines in a precise way the amount of entanglement used in the protocol. Therefore, finding the smallest group is crucial, a task that is likely to be computationally intensive in many cases. We have presented possible ways to find this smallest group but leave as an open problem the question of optimizing this search.

The protocol considered in this paper is just one that has been presented in \cite{OurNLU}, and it is possible to identify the others as special cases of this one. If, for example, the $U(f)$ all commute with each other, then $\UC$ is a controlled-unitary \cite{OurNLU}, which can be written in the form
\begin{eqnarray}
	\UC=\sum_j P_j\otimes \VC_j,
\end{eqnarray}
with $\VC_j$ an arbitrary set of unitaries on $\HC_B$ and the $P_j$ are projectors on $\HC_A$ satisfying $\sum_j P_j=I_A$. This expansion as a controlled-unitary may be found directly from the expansion in terms of the $U(f)$ by simply diagonalizing the latter and identifying the $P_j$ as projectors onto the subspaces corresponding to the individual one-dimensional irreps. Note that some of these irreps may be repeated in the $U(f)$ representation, and this will lead to higher-rank projectors $P_j$ \cite{OurNLU}.

The protocol of \cite{OurNLU} for controlled-unitaries may have advantages over the protocol discussed here, as the local gates used by Alice and Bob may be easier to implement in practice (the unitary $M$ of Figure~\ref{fig:bipartM} is replaced by a simpler set of gates). This may also be true of the protocol for what has been called a ``double unitary" representation in \cite{OurNLU}, with an expansion of the form
\begin{eqnarray}
	\UC =\sum_{f\in G}c(f)U(f)\otimes V(f),
\end{eqnarray}
where $\{U(f)\}$ and $\{V(f)\}$ are each a projective representation of the same group $G$ and $c(f)$ are complex coefficients. If the set of operators $W(f)$, found from (\ref{eq:W}), are proportional to unitaries that represent $G$, then we have this type of double unitary representation. However, if the $U(f)$ are linearly dependent, then even if the $W(f)$ are not proportional to unitaries, there may be another set that can be used that are. Finding the easiest way to determine if this is the case is left for future study.

\section{Acknowledgments}
This work has been supported in part by the National Science Foundation through Grants PHY-0456951 and PHY-0757251, as well as by a grant from the Research Corporation. In addition, I would like to thank Li Yu for numerous valuable discussions, and to acknowledge the gracious hospitality of the Department of Physics at Lewis and Clark College where this paper was written.

\appendix
\section{Proof of theorem~\ref{th:block}}\label{app:block}
Here I prove that if the $A_k$ are the expansion coefficients on one side for a non-singular bipartite operator, $\MC=\sum_jA_j\otimes B_j$, then there always exists a single, fixed unitary $V$ such that the finest block diagonal form of the $V^\dagger A_k$ is the same as that of the $A_j^\dagger A_k$. The argument provides a direct method of constructing $V$.

Given that $A_j^\dagger A_k$ are block diagonal for every $j,k$ (assume, without loss of generality, we work in the appropriate basis to bring them to this form), with blocks of size $n_1$, $n_2$, etc., in that order. Then this means the following: The first $n_1$ columns of $A_k$ are orthogonal to columns $n_1+1, n_1+2,\cdots, D_A$ of every $A_j$, including $A_k$; columns $n_1+1, \cdots, n_1+n_2$ of $A_k$ are orthogonal to columns $n_1+n_2+1,  \cdots, D_A$ of every $A_j$ (and also to the first $n_1$ columns, of course); etc. So we can partition all the columns of all the $A_k$ into subsets of vectors, where any vector in one set is orthogonal to every vector in any other subset.

As ${\MC}$ is non-singular, the collection of all these vectors spans ${\cal H}_A$ (Alice's Hilbert space). If this were not the case, then there would exist a state $|a\rangle\in\HC_A$ with the property that for any $|b\rangle\in\HC_B$, the ket $|\xi\rangle=|a\rangle\otimes|b\rangle$ lies in the nullspace of $\MC^\dagger$, contradicting the assumed non-singularity of $\MC$. Furthermore, we have the stronger statement,

\begin{lemma_block}
For any subset of the integers $1$ to $D_A$, call this subset $S$ having cardinality $n$, collect all the vectors forming columns $m_1, m_2,  \cdots, m_n$, with $m_k \in S$, from all the $A_j$ taken from an expansion of a non-singular (bipartite) matrix ${\MC}=\sum_jA_j\otimes B_j$. Then, this collection of vectors spans a space of dimension at least $n$.
\end{lemma_block}
Proof of lemma:  Defining states $|1\rangle,  \cdots, |D_A\rangle$ as the standard basis of ${\cal H}_A$ (the subsystem upon which the $A_j$ act), we have for example, that $A_j$ maps state $|m_1\rangle$ to the vector defined by column $m_1$ of $A_j$. If the statement of the lemma were false, than the collection of vectors $\{|m_k\rangle|l\rangle\}$ (where $k=1, \cdots,n$;  $l=1, \cdots,D_B$; and $|l\rangle$ are the standard basis of ${\cal H}_B$), which together span a subspace of dimension $nD_B$, would be mapped by ${\MC}$ into a subspace of dimension strictly less than $nD_B$. This is a contradiction, since a non-singular matrix maps subspaces into subspaces of equal dimension, proving the lemma. \hspace{\stretch{1}}$\blacksquare$

Due to this lemma, we see that the partitioning mentioned earlier is such that every subset of vectors spans a subspace of dimension equal to the number of columns it is drawn from ($n_\alpha$ for the $\alpha^{th}$ subset). No subset can span a larger space, since this would mean, considering all the subsets together, having more than $D_A$ orthogonal vectors on ${\cal H}_A$. Choose a basis for each of these subspaces, call these bases $\BC_1$ for the subset of the first $n_1$ columns, $\BC_2$ for that of the second $n_2$ columns, etc. Then we may conclude that every $A_k$ maps $\{|1\rangle, \cdots,|n_1\rangle\}$ into $\BC_1$, $\{|n_1+1\rangle, \cdots,|n_1+n_2\rangle\}$ into $\BC_2$, and so on. Define $V$ to be a unitary that mirrors these mappings. Then, $V^\dagger$ undoes the action of each of the $A_k$ on these subsets of basis states, mapping $\BC_1$ onto $\{|1\rangle, \cdots,|n_1\rangle\}$, $\BC_2$ onto $\{|n_1+1\rangle, \cdots,|n_1+n_2\rangle\}$, etc., and $V^\dagger A_k$ maps $\{|1\rangle, \cdots,|n_1\rangle\}$ to $\{|1\rangle, \cdots,|n_1\rangle\}$, $\{|n_1+1\rangle, \cdots,|n_1+n_2\rangle\}$ to $\{|n_1+1\rangle, \cdots,|n_1+n_2\rangle\}$, and so on. This means that $V^\dagger A_k$ is block diagonal in the same form as all the $A_j^\dagger A_k$ and we are done, including having constructed the unitary $V$ (as the above argument indicates, the choice of $V$ is generally not unique). Note also that these arguments imply these two sets of operators, $\{V^\dagger A_k\}$ and $\{A_j^\dagger A_k\}$, take on their respective finest block diagonal forms in one and the same basis.

\section{Searching for projective representations}\label{app:proj}
We wish to extend our search for the smallest group to include projective representations of a group $G$ \cite{Kim_group}. If for each admissible $N$, no group of this order exists that has the desired ordinary irreps, we then need to determine if any such group exists that instead has a set of projective irreps matching the required dimensions. Note the additional requirement that all irreps must share the same factor system, since if the representation by $U(f)$ is decomposed into a direct sum of irrep blocks, then (\ref{muFG}) will be satisfied if and only if it is satisfied block-by-block with a fixed $\mu(f,g)$. This point indicates that if $d_\lambda=1$ for any $\lambda$, then it must be that the representation is an ordinary one, since one-dimensional projective irreps are always equivalent to ordinary irreps.

Below, we will describe one possible method of searching for a group having the desired irrep dimensions, which will utilize the following lemma.
\begin{pirrep}
	A projective irrep of dimension $d$ exists for a group $G$ if and only if an ordinary irrep of that dimension exists for a central extension $L$ of $G=L/K$ by some cyclic group $K$ of order $r$, where $r$ divides $|G|$. Furthermore, all projective irreps of $G$ that share the same factor system $\{\mu(f,g)\}$ may be derived from ordinary irreps of a single such extension, $L$.
\end{pirrep}
Proof: The ``if" part follows directly from the well-known relationship between central extensions and projective representations \cite{Kim_group}. To prove the converse, suppose a projective irrep $\widetilde\Gamma$ of dimension $d$ and factor system $\widetilde\mu$ exists. Then there also exists a projective irrep $\Gamma$ of this dimension having a normalized factor system $\mu$ \cite{Kim_group}. That is, for all $f$ and $g$, 
\begin{eqnarray}
	\Gamma(f)\Gamma(g)=\mu(f,g)\Gamma(fg),
\end{eqnarray}
where by normalized we mean that $\mu(f,g)=\omega_r^{n(f,g)}$ with $\omega_r=e^{2\pi i/r}$ for some $r$ dividing $|G|$ and $n(f,g)$ is an integer. Consider an extension $L$ of $G$ with elements $(m,g) \in L$, where $g\in G$, $m$ is an integer lying between $0$ and $r-1$, and group multiplication is defined as 
\begin{eqnarray}
	(l,f)(m,g)=(l+m+n(f,g),fg),
\end{eqnarray}
with addition mod $r$. Given our definition of $n(f,g)$ in terms of the normalized factor system $\mu(f,g)$, it is not difficult to show that this multiplication satisfies the properties of a group. 

Next, construct from the irrep $\Gamma(g)$ the function
\begin{eqnarray}
	\DC(m,g)=\omega_r^m\Gamma(g),
\end{eqnarray}
which satisfies
\begin{eqnarray}
	\DC(l,f)\DC(m,g)=\omega_r^{l+m}\Gamma(f)\Gamma(g)=\omega_r^{l+m+n(f,g)}\Gamma(fg)=\DC(l+m+n(f,g),fg),
\end{eqnarray}
so is clearly an ordinary irrep of $L$ having dimension $d$, and it is readily seen that $L$ is a central extension of $G$ by the cyclic group of order $r$. Furthermore, we see that $\mu(f,g)$ defines $n(f,g)$ which in turn, along with the given group multiplication defined for $G$, defines the group multiplication for $L$. Therefore, these arguments show that every projective irrep $\Gamma$ of $G$ that shares this factor system $\mu$ gives rise to an ordinary irrep $\DC$ of this single extension $L$, completing the proof. \hspace{\stretch{1}}$\blacksquare$

One way to determine all the projective irreps of a group $G$ is to find the Schur cover or representation group of $G$ \cite{Kim_group}, but this is generally a difficult task and in any case, provides much more information than is needed here. Let us now describe another possibility, which may be more efficient. We seek a group $G$ of order $N$ with projective irreps of specific dimensions $\{d_\lambda\}$, all sharing a single factor system. By the previous lemma, we know that each such irrep exists iff an ordinary irrep of that dimension exists for a central extension $L$ satisfying the stated conditions of the lemma. So, we can consider groups of order $rN$ where $r$ divides $N$, and check if one such group has ordinary irreps with dimensions matching one or more of the $d_\lambda$. If we find such an $L$, and if $L$ has a cyclic subgroup $K$ of order $r$ in its center, then $G=L/K$ has a projective irrep of dimension $d_\lambda$ (note that $L$ will have a cyclic subgroup of order $r$ in its center iff this irrep includes characters equal to $d_\lambda,~\omega_r d_\lambda,~\omega_r^2 d_\lambda,\cdots,~\omega_r^{r-1} d_\lambda$). If, after searching groups of order $rN$ for all $r$ dividing $N$, we have not found all the required irrep dimensions, then no group of order $N$ can be used for the expansion of $\UC$, and one should move to groups of larger order. On the other hand, if all required irrep dimensions are found, then one must check to make sure that for each central extension $L_\lambda$ (providing the projective irrep of dimension $d_\lambda$), the group $G=L_\lambda/K_\lambda$ is the same for all $\lambda$. If it is, one more check is necessary, that one can choose projective irreps of the required dimensions that all share the same factor system for that particular group $G$. If so, then we may use the group $G$ to expand $\UC$. 

Alternatively, since the lemma guarantees a single $L$ providing all the irreps that share a single factor system for a given $G$, we can simply search for an $L$ of order $rN$ with irrep dimensions matching the desired set. In addition, noting that every normalized factor system may be written as powers of $\omega_{N}$ ($r=N$), one may choose to restrict the search to groups of order $N^2$ since if the required set of projective irreps sharing a single factor system does in fact exist, there must be an extension group of this order providing all these irreps.

The simplest example is for $N=4$ when we need only one irrep of dimension $d_1=2$. Since every group of order $|G|<6$ is Abelian, all such groups have only one-dimensional ordinary irreps. However, setting $r=4$ so that $\omega_r=i$ and looking at groups of order $rN=16$, we find the group ($L$) generated by the Pauli matrices having a two-dimensional irrep with characters $2,2i,-2,~\textrm{and}~-2i$, which tells us that the center of this group is isomorphic to the cyclic group, $C_4$, of order 4. In addition, $L/C_4=C_2 \times C_2\equiv G$. A valid two-dimensional projective irrep for $G$ consists of the identity along with the Pauli matrices, and may be readily obtained from the two-dimensional ordinary irrep of $L$. Note that one can also construct a projective representation using $r=2$, the central extension then being $Q8$ or $D4$ (either will work).



\begin{thebibliography}{10}
\bibitem{HermEqBlocks}
This sort of thing is more easily determined if we instead consider block-diagonalizing the Hermitian combinations, $A_j^\dagger A_k+A_k^\dagger A_j$ and $i(A_j^\dagger A_k-A_k^\dagger A_j)$ --- this does not alter any conclusions --- and compare eigenvalues of the individual blocks. If for any pair of blocks, these eigenvalues match for every $j,k$, one can then check if a fixed $\TC$ will work for all $j,k$ by looking at the corresponding eigenvectors for different $j,k$ pairs.
\end{thebibliography}

\end{document}